\documentclass[11pt]{article} 
\usepackage{hyperref} 
\pdfoutput=1 
\usepackage{graphicx}
 
\begin{document} 
 
\title{Faraday instability on a network} 
 
\author{G. Delon, D. Terwagne, N. Vandewalle, S. Dorbolo and H. Caps
\\\vspace{6pt} GRASP, Opto-fluidics -- Physics Department B5 \\
Universit\'e de Li\`ege \\
B-4000 Li\`ege -- Belgium
}

\maketitle

\begin{abstract}
Faraday waves are generated at the air/liquid interface inside an array of square cells.  As the free surface inside each cell is destabilizing due to the oscillations, the shape of the free surface is drastically changing. Depending on the value of the frequency $f$ of oscillations, different patterns are observed inside each cell. For well defined $f$ values, neighboring cells are observed to interact and a general organization is noticed. In such a situation, initially disordered structures lead to a general pattern covering the entire liquid pool and a spatial order appears all over the cell array. This abstract is related to a fluid dynamics video for the gallery of fluid motion 2009. 

\end{abstract} 
 

Under periodic oscillations, an air/liquid interface may become unstable to surface waves. Theses waves are known as Faraday waves, after the pioneer work of Faraday \cite{faraday}. As a parametric instability, different kind of wave patterns may appear, depending on the excitation parameters \cite{douady,muller}. It has also been noticed that the pinning of the contact line at the wall/liquid interface plays a role in the dynamics of these waves \cite{pinning}.

\

In the present experiments, a liquid pool is divided into separated square cells by mean of a plexiglass grid. The walls separating the cells are $2.5$~cm in height and have a width of $1$~mm.  These cells are partially filled with water or $10$~cSt silicon oil. The vertical oscillations are characterized by a frequency $f$ ranging from $10$~Hz to $30$~Hz and an amplitude of typically $1$~mm. With the help of a high speed video camera, the emergence of Faraday waves is investigated. Beside individual behaviors the different cells, collective effects have been observed. In the linked videos, two collective modes are presented. The first one appears at a frequency of $10$~Hz and an acceleration above the threshold for Faraday instability. In this mode, large liquid bumps are generated at the cell intersections. These bumps are located on the dual network. This means that bumps are apart from two cells from their first neighbor. Two dual networks of bumps are observed to alternate, each one appearing at the Faraday frequency ($f/2$). This mode is called `dual mode'.

In the second mode, when the frequency is $f=14$~Hz, diagonal waves are observed to be generated inside each cell. At the cell intersections, these waves interact and generate bumps at these points. Taking a diagonal over all the grid, leads to a series of bumps. The bumps are thus placed in quincunx at the cell intersections. This time, bumps are apart from each others from only one cell diagonal. They alternate at the Faraday frequency ($f/2$). This mode has been called `diagonal mode'.

 \begin{figure} [h]
\begin{center}
\includegraphics[height=4.85cm]{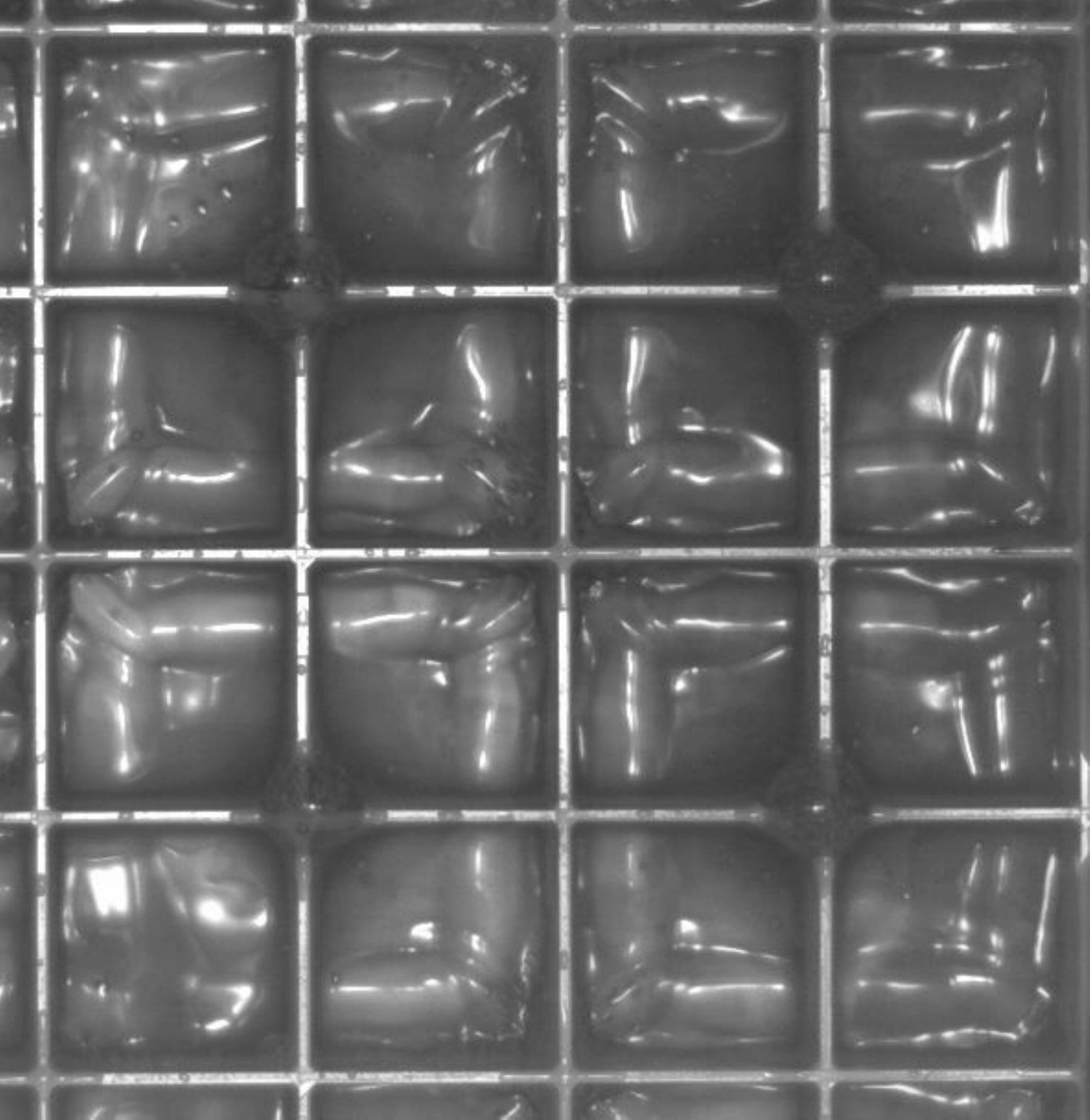}\ \includegraphics[height=4.85cm]{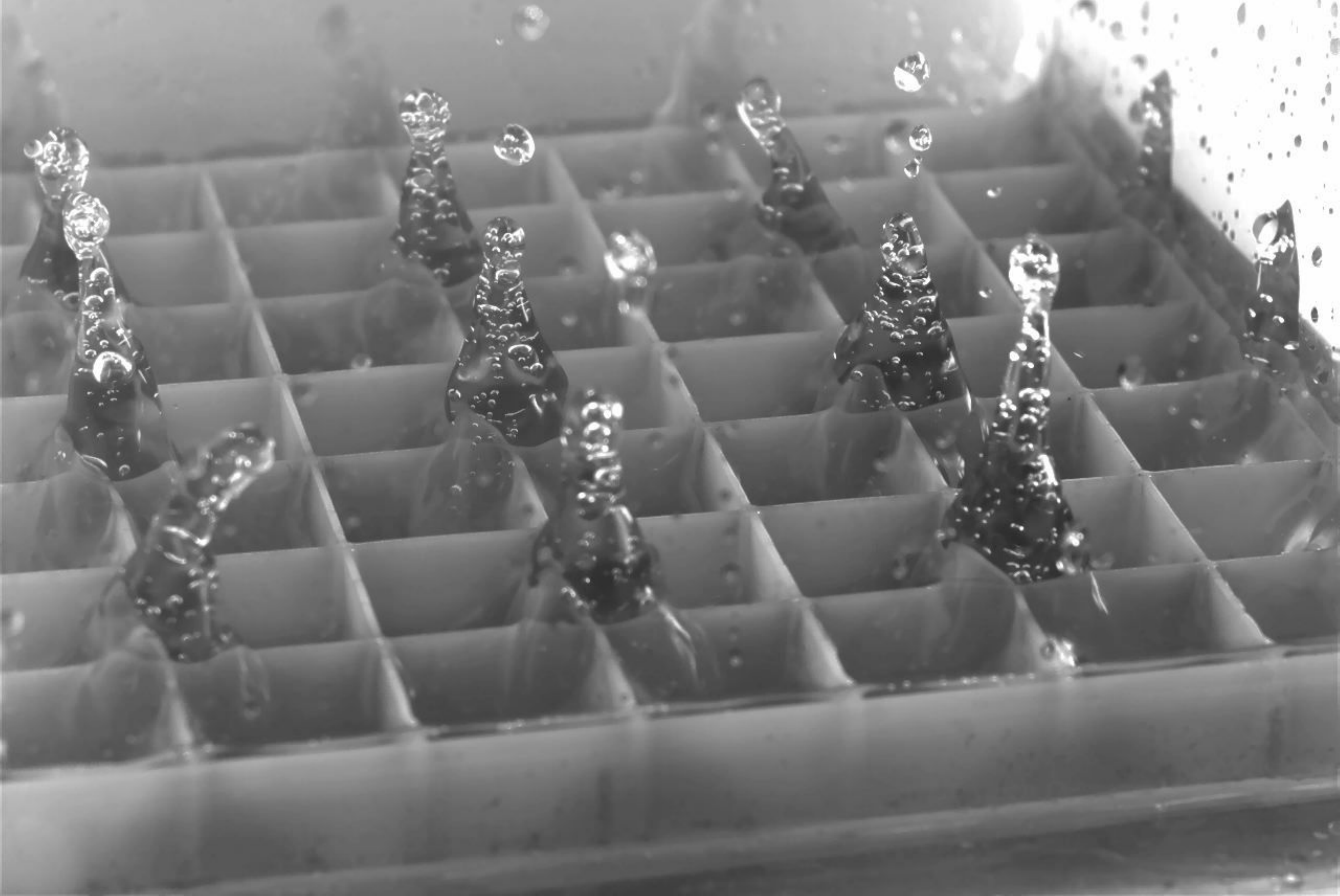}
\includegraphics[height=5cm]{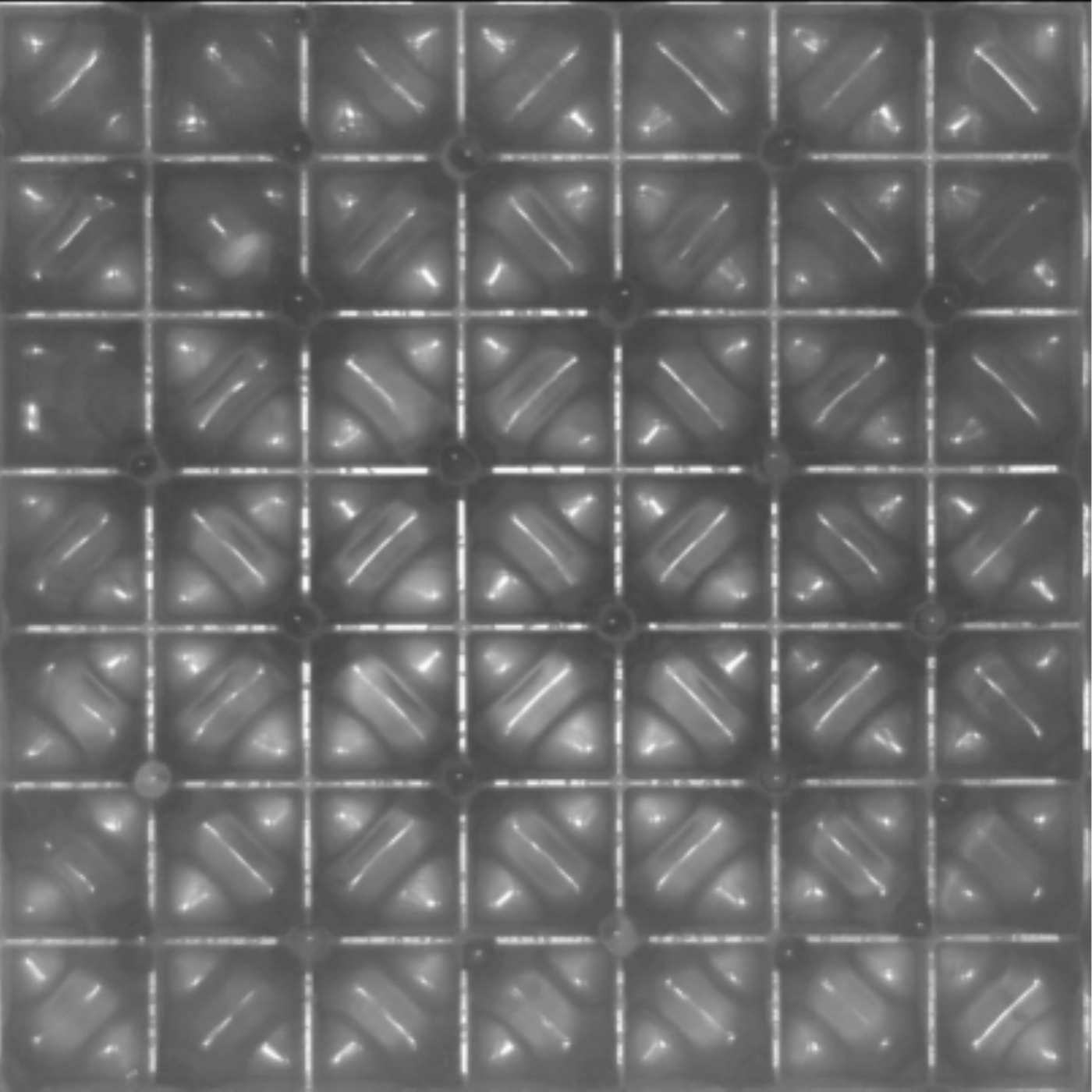}\ \includegraphics[height=5cm]{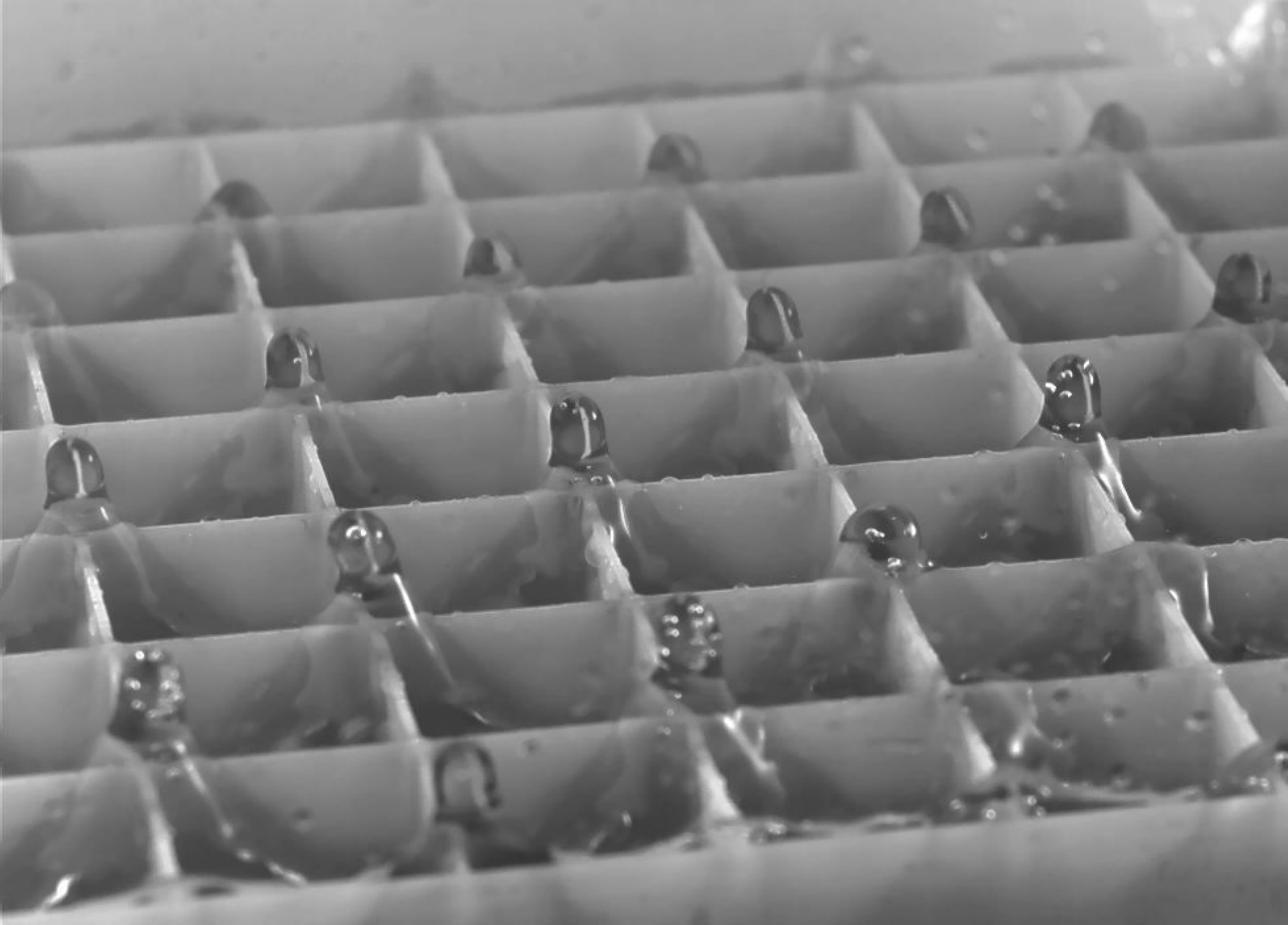}
\end{center}
\caption{(Top line) Top and side views of dual mode at $f=10$~Hz. Bumps are observed on the dual network, separated by two cells from each others. (Bottom line) Top and side views of the diagonal mode at $f=14$~Hz. Bumps are in quincunx at the cell intersections. They are apart from one cell diagonal. }
\end{figure}

\

\noindent G. Delon thanks Belspo for financial support. S.Dorbolo is financially supported by FNRS-FRS Belgium.
 
\

\noindent Videos can be found with the following links : 
\begin{itemize}
\item \href{http://ecommons.library.cornell.edu/bitstream/1813/14069/3/delon_mpeg1.mpg}{Video 1 - Low res.} and 
\item \href{http://ecommons.library.cornell.edu/bitstream/1813/14069/2/delon_mpeg2.mpg}{Video 2 - High res.}. 
 \end{itemize}

%
\end{document}